\documentclass[runningheads]{cl2emult}

\usepackage{makeidx}  
\usepackage{graphicx} 
\usepackage{subeqnar} 
\usepackage{multicol} 
\usepackage{cropmark} 
\usepackage{lnp}      
\makeindex            

\begin{document}
\title*{FIRBACK Source Counts and Cosmological Implications}
\toctitle{FIRBACK Source Counts Cosmological Implications}
%
%
\titlerunning{FIRBACK Source Counts and Cosmological Implications}
%
\author{H.~Dole\inst{1}
\and R.~Gispert\inst{1}
\and G.~Lagache\inst{1}
\and J-L.~Puget\inst{1}
\and H. Aussel\inst{2,3}
\and F.R. Bouchet\inst{4}
\and P. Ciliegi\inst{5}
\and D.L. Clements\inst{6}
\and C.J. Cesarsky\inst{7}
\and F.X. D\'esert\inst{8}
\and D. Elbaz\inst{2}
\and A. Franceschini\inst{3}
\and B. Guiderdoni\inst{4}
\and M. Harwit\inst{9}
\and R. Laureijs\inst{10}
\and D. Lemke\inst{11}
\and R. McMahon\inst{12}
\and A.F.M. Moorwood\inst{7}
\and S. Oliver\inst{13}
\and W.T. Reach\inst{14}
\and M. Rowan-Robinson\inst{13}
\and M. Stickel\inst{11}
}
\authorrunning{Herv\'e Dole \it{et al.}}
%
%
\institute{Institut d'Astrophysique Spatiale, Orsay, France
\and Service d'Astrophysique, CEA/DSM/DAPNIA Saclay, France
\and Osservatorio Astronomico di Padova, Italy
\and Institut d'Astrophysique de Paris, France
\and Osservatorio Astronomico di Bologna, Italy
\and Cardiff University, UK
\and ESO, Garching, Germany
\and Laboratoire d'Astrophysique, Observatoire de Grenoble, France
\and 511 H.Street S.W., Washington, DC 20024-2725
\and ISOC ESA, VILSPA, Madrid, Spain
\and MPIA, Heidelberg, Germany
\and Institute for Astronomy, University of Cambridge, UK
\and Imperial College, London, UK
\and IPAC, Pasadena, CA, USA
}

\maketitle              
\vspace{-0.4cm}
\begin{abstract}
FIRBACK is a one of the deepest surveys performed at 170 $\mu m$ with ISOPHOT onboard
ISO, and is aimed at the study of cosmic far infrared background sources. About 300 galaxies are
detected in an area of four square degrees, and source counts present a strong slope of 2.2 on an
integral "logN-logS" plot, which cannot be due to cosmological evolution if no K-correction is
present. The resolved sources account for less than 10\% of the Cosmic Infrared Background at
170 $\mu m$.
In order to understand the nature of the sources contributing to the CIB, and to explain deep
source counts at other wavelengths, we have developed a phenomenological model, which
constrains in a simple way the luminosity function evolution with redshift, and fits all the
existing deep source counts from the mid-infrared to the submillimetre range.
\end{abstract}
\vspace{-0.8cm}
%
\section{Introduction}
The Cosmic Infrared Background (CIB), due to the accumulation of galaxy emission at all
redshifts along the line of sight in an instrument beam, is a powerful tool for studying galaxy
evolution. FIRBACK (\cite{puget99a}, \cite{dole99}), one of the deepest surveys performed at
$170\,\mu m$, is aimed at the study of the CIB, in two complementary ways:
\begin{itemize}
\vspace{-0.2cm}
\item study the resolved sources (this paper)
\item study the background fluctuations (\cite{lagache99a}, \cite{lagache99b})
\end{itemize}
\vspace{-0.2cm}
Throughout this paper we use a cosmology with $h=0.65$, $\Omega=1$ and $\Lambda=0$.
%
\section{The FIRBACK Survey}
FIRBACK, is a survey of 4 square degrees in 3 high galactic latitude fields, chosen to have as
low an HI column-density as possible, typically \- $N_H \simeq 10^{20} cm^{-2}$, and if
possible multiwavelength coverage.
Observations were carried with ESA's Infrared Space Observatory (ISO, \cite{kessler96}) with
the ISO\-PHOT photometer \cite{lemke96} in raster mode (AOT P22) with the \verb+C200+
camera and \verb+C_160+ broadband filter centered at $\lambda = 170\, \mu m$. 
A detailed description of the reduction, data processing, and calibration will be discussed in
\cite{lagache2000}, whereas the analysis of the complete survey will be discussed in
\cite{dole2000}.
\section{Source Counts at $170\,\mu m$}
\begin{figure}
\centering
\includegraphics[width=.85\textwidth]{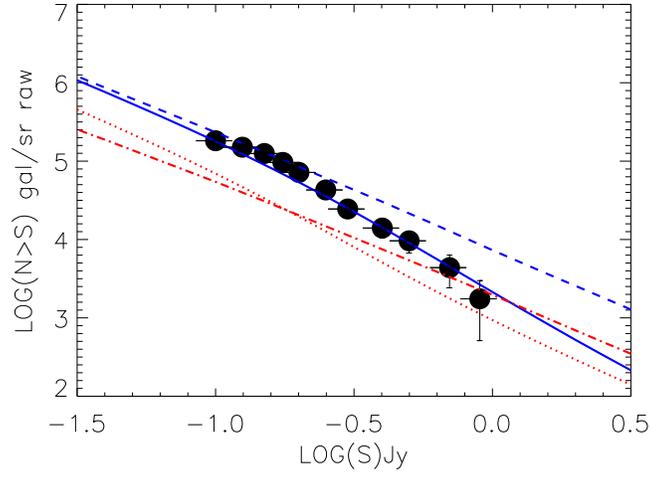}
\vspace{-0.3cm}
\caption[]{FIRBACK integral source counts at $170\,\mu m$ not corrected for incompleteness.
Models from \cite{guiderdoni98}: A (dot) with evolution, E (solid) with evolution + ULIRGs.
Models from \cite{franceschini98}: without evolution (dot-dash), with evolution (dash).}
\label{firback_count}
\end{figure}

Preliminary FIRBACK integral source counts at $170\,\mu m$ (Fig.~\ref{firback_count}), not
corrected for incompleteness, show a strong slope of 2.2 between 120 and 500 mJy. This strong
slope is not explained by a K-correction or cosmological evolution alone: both must be present;
the K-correction is the ratio, at a given wavelength, of the emitted flux over the redshifted flux.\\
Non (or low) evolution scenarii, or extrapolation of IRAS counts, are not able to reproduce the
observed counts. For illustration, we plot in Fig.~\ref{firback_count} the non-evolution model
from \cite{franceschini98} and the evolution model A from \cite{guiderdoni98}.
On the other hand, evolutionary models from \cite{franceschini98} and from \cite{guiderdoni98}
(model E: evolution + ULIRGs) give a better agreement.
At this wavelength, FIRBACK sources account for only 3\% of the background.
%
\section{Modelling the Evolution of Galaxies}
%
\subsection{Method}
\begin{figure}
\centering
\includegraphics[width=.85\textwidth]{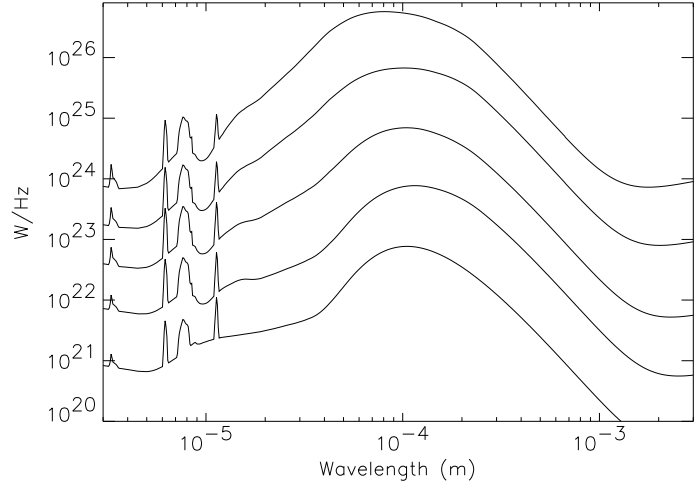}
\vspace{-0.2cm}
\caption[]{Template spectra used for simulating source counts; from bottom to top: $L =
10^{9},\, 10^{10},\, 10^{11},\, 10^{12},\, 10^{13}\, L_{\odot}$.}
\label{template_spectra}
\end{figure}

Our philosophy is to make a phenomenological model that explains all the observed deep
source counts and reproduces the CIB in the mid-IR to submillimetre range.
One way to do this is to constrain the evolution with redshift of the luminosity function (LF) in
the infrared, given:
\begin{itemize}
%
%
\item templates of galaxy spectra
\item the energy density available at each redshift
\end{itemize}
For the first point, we used template galaxy spectra based on IRAS colors \cite{maffei94}
modified to account for recent ISO observations, in particular the absorption feature near
$10\,\mu m$ at high luminosity; PAH features are present in the mid-infrared \cite{desert90},
even if their strength seems larger than the observations: this is not a problem because the right
amount of energy is present in each peak, and we convolve the spectrum with the filter spectral
response (Fig.~\ref{template_spectra}).
For the second point, we used the inversion of the CIB spectrum by \cite{gispert2000} and
\cite{puget99b}, which gives with good accuracy the energy density available at redshifts
between 1 and 3. This energy density is the integral of the luminosity function at each redshift,
but there is no unique solution for the LF shape.
\vspace{-0.3cm}
%
\subsection{Evolving Luminosity Function}
\begin{figure}
\centering
\includegraphics[width=0.95\textwidth]{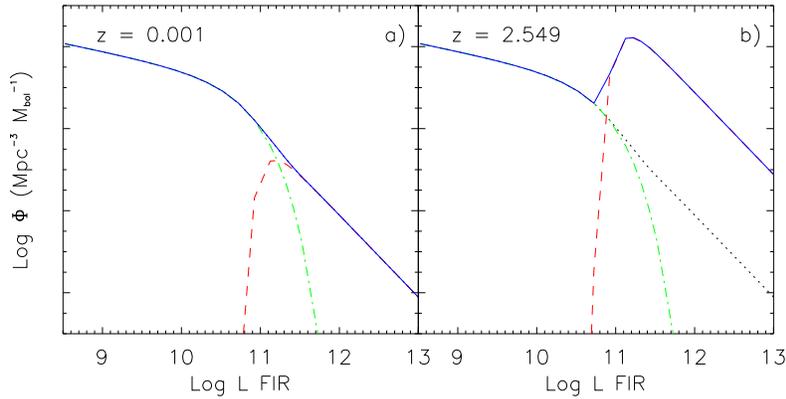}
\vspace{-2.0cm}
\caption[]{a: Luminosity Function at z=0 (solid line); normal galaxy (dot-dash); ULIRG
(dash-dash). b: Luminosity Function at z=2.5 (solid line); normal galaxy (dot-dash) ULIRG
(dash-dash) and local LF (dots).}
\label{lf_z0}
\end{figure}

The question is: how to evolve the LF with redshift ? \\
Fig.~\ref{lf_z0} represents the LF of \cite{sanders96} at a redshift z=0, renormalized to
$h=0.65$ and having the same integral as \cite{soifer91}. Our constraint on the LF redshift
variation is the energy density available at each redshift \cite{gispert2000}. Usually, authors
apply a pure density evolution to the LF as a  function of redshift (i.e. a vertical shift), or a pure
luminosity evolution (a horizontal shift).  This does not work for FIRBACK source counts: There
is no alternative other than adding the evolution to one part of the LF only. This part is
constrained by IR and submm observations: this is the bright end of the LF.
Fig.~\ref{lf_z0} represents our decomposition of the local LF into two parts:
\begin{itemize}
\vspace{-0.2cm}
\item left part: ``normal'' galaxies
\item right part: ULIRG's, centered on a luminosity $L_{ULIRG} \simeq 2.0 \times 10^{11}
L_{\odot}$, where $L_{ULIRG}$ is the free parameter; we get the same value as \cite{tan99}
\end{itemize}
In our model, "without evolution" means that the local LF (Fig.~\ref{lf_z0}a) is taken, and is the
same at every redshift.
In the evolutionary scenario, only the ULIRG part moves, in such a way that the integral of the
LF equals the constraint given by the CIB inversion at each redshift. The maximum is reached at
$z \simeq 2.5$, and Fig.~\ref{lf_z0}b represents the LF at this redshift. We neglect at this stage
the evolution of ``normal'' galaxies, which do not play a crucial role in the mid-infrared to
submillimetre wavelength range.
%
%
\vspace{-0.2cm}
\subsection{Model of Source Counts at $170\,\mu m$}
\begin{figure}
\centering
\includegraphics[width=.85\textwidth]{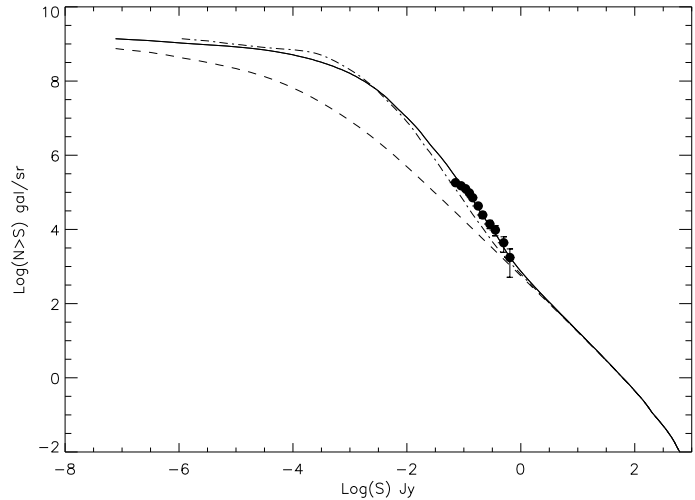}
\vspace{-0.3cm}
\caption[]{Observed Counts and Models at $170\,\mu m$, with (solid line) and without (dashed
line) evolution.  The model with evolution and without K-correction is the dot-dash line.}
\label{create_counts170}
\end{figure}

The model at $170\,\mu m$, together with our data, is presented in Fig.~\ref{create_counts170}.
The brightest point of the observed counts is compatible with all our models, in particular the
non-evolutionary scenario, which is expected for local sources.  We also show the effect of the
K-correction, which steepens the integral source count slope. Our evolutionary scenario fits the
data within the error bars. Most of the background is expected to be resolved into sources once
we are able to detect sources at the mJy level. This wavelength region, in which evolutionary
effects are particularly important and where there are good prospects for detecting higher redshift
sources because of the K-correction, is nowadays probably the best-suited range for probing
galaxy evolution from space in the far-infrared. 
%
\vspace{-1.0cm}
%
\subsection{Models of Source Counts at other wavelengths}
\begin{figure}
\centering
\includegraphics[width=.85\textwidth]{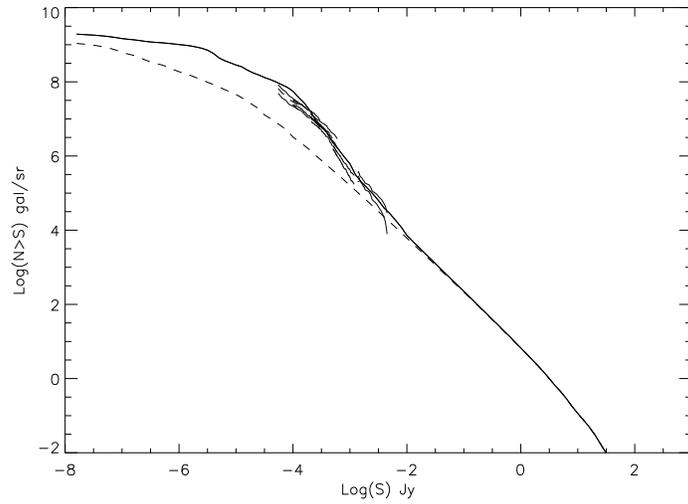}
\vspace{-0.3cm}
\caption[]{Models at $15\,\mu m$ with (solid line) and without (dashed line) evolution, and
observed counts from Elbaz et al. \cite{elbaz99} (thin lines).}
\label{create_counts15}
\end{figure}

\begin{figure}
\centering
\includegraphics[width=.85\textwidth]{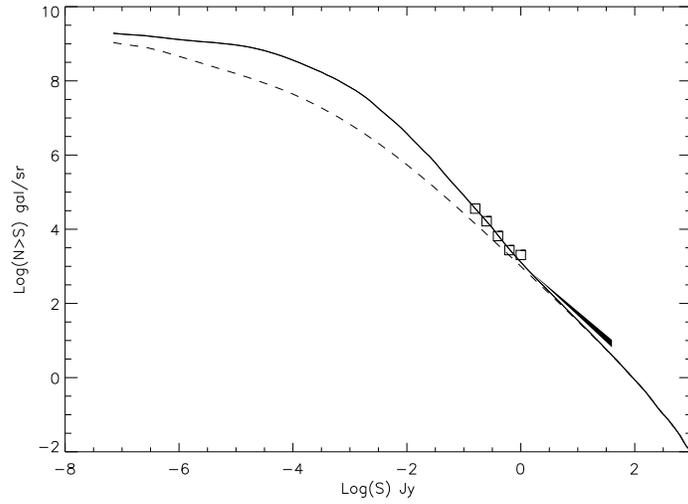}
\vspace{-0.3cm}
\caption[]{Models at $90\,\mu m$ with (solid line) and without (dashed line) evolution, and
observed counts from Efstathiou et al. \cite{efstathiou99} (squares) and IRAS counts
\cite{efstathiou99} (solid area)}
\label{create_counts90}
\end{figure}

\begin{figure}
\centering
\includegraphics[width=.85\textwidth]{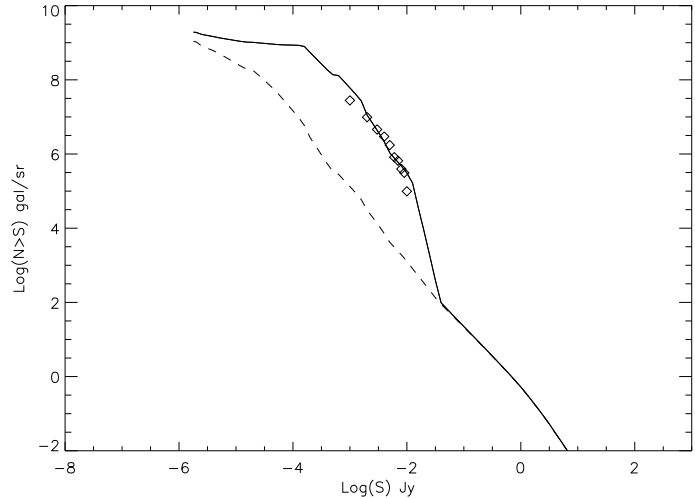}
\vspace{-0.3cm}
\caption[]{Models at $850\,\mu m$ with (solid line) and without (dashed line) evolution, and
observed counts from Barger et al. \cite{barger99} (diamonds).}
\label{create_counts850}
\end{figure}

Data at $15\,\mu m$ \cite{elbaz99}, $90\,\mu m$ \cite{efstathiou99} and $850\,\mu m$
\cite{barger99}, with our models are presented
in Fig.~\ref{create_counts15}, Fig.~\ref{create_counts90} and Fig.~\ref{create_counts850}
respectively.\\
At $15\,\mu m$, both the slopes and the little ``waves'' in the counts appear in the models with
the combined effects of the K-correction and the evolution. At $90\,\mu m$, the K-correction
does not emphasize the differences between the scenarii of evolution or non-evolution. At
$850\,\mu m$, the ``non smooth'' appearance of our model is due to the discretization of the LF.
%
%
%
%
\section{Discussion}
%
\subsection{The bright end luminosity function evolution model}
Our model of the evolving LF scenario fits most of the existing deep survey data from space (mid
and far infrared) and ground (submillimetre).
It is also compatible with the observational estimate of the LF in \cite{lilly99}.
The strong observational constrain of multiwavelength source counts is thus explainable by a
simple evolutionary law of the LF: 
the Bright End Luminosity Function Evolution (BELFE) model. 
One simple model is in agreement with all up-to-date observables and reproduces the
background.
%
\vspace{-0.2cm}
\subsection{Redshift and Nature of the sources}
Another crucial observational test for our
model is the predicted vs observed redshift distribution. Although the statistics are poor, we have
some evidence that most of the ISOCAM sources lie at redshift between 0 and 1.4 with a median
at 0.8 \cite{aussel99}, and that most of the SCUBA sources lie at redshifts greater than 2
\cite{barger99}.
Our predicted redshift distributions at these two wavelengths are in agreement with the existing
observations.\\
What about FIRBACK $170\,\mu m$ sources ? Our predicted redshift distribution shows that
most of the sources lie at redshifts below 1.5, with a median comparable to that of the ISOCAM
sources.  This means that we are sensitive both to local sources and sources beyond redshift $z
=1$.  All the FIRBACK sources with known redshifts (less than 10) \cite{dennefeld99},
\cite{scott2000} are in this range. Two submillimetre sources are at $z > 1$, and a few visible
sources are at $z \simeq 0.2$.\\
It is difficult to address the question differences between PHOT and CAM sources, because their
redshift distributions (both expected and observed) are similar.  About half of the $170\,\mu m$
sources have $15\,\mu m$ counterparts.
%
\vspace{-0.3cm}
\section{Conclusion}
We have presented results from the FIRBACK survey, one of the largest ISO programs, dealing
with resolved sources of the CIB: our source counts at $170\,\mu m$ show strong evolution. This
evolution is explained by a simple law involving the redshift of the luminosity function, which
differs from pure density or luminosity evolution: the 
``Bright End Luminosity Function Evolution'' model.
The model fits all the existing source counts at 15, 90, 170 and 850 $\mu m$, and also predicts a
redshift distribution in agreement with the (sometimes sparse) observations. This powerful tool,
based on observational constraints on the CIB spectrum inversion and the local Luminosity
Function, and on assumed template galaxy spectra, not only agrees with existing data but also is
able to make useful predictions on source counts or CIB fluctuations.  These predictions may be
useful for planning the utilization of major telescopes of the future, such as SIRTF, Planck,
FIRST and ALMA.
All the FIRBACK materials are available at: \verb|http://wwwfirback.ias.fr|.
%

\clearpage
\addcontentsline{toc}{section}{Index}
\flushbottom
\printindex

\end{document}